\documentclass[a4paper, 12 pt, onecolumn]{ieeeconf}

\usepackage{cite}
\usepackage{amsmath,amssymb,amsfonts}
\usepackage{algorithm}
\usepackage{algorithmic}
\usepackage{graphicx}
\usepackage{textcomp}
\usepackage{xcolor}

\usepackage{courier}
\usepackage{enumerate}
\usepackage{stmaryrd}
\usepackage{subfigure}

\usepackage{moresize}
\usepackage{hyperref}
\usepackage{cleveref}

 \usepackage{lipsum}

\usepackage{listings}
\usepackage{multirow}
\usepackage[all,cmtip]{xy}
\usepackage{tikz}
\usetikzlibrary{automata, positioning, arrows.meta, decorations.markings}
\definecolor{improve}{gray}{0.5}
\definecolor{tikz.lightgrey}{rgb}{0.6, 0.6, 0.6}
\definecolor{tikz.red}{RGB}{190, 30, 45}
\definecolor{tikz.green}{RGB}{0, 147, 68}
\definecolor{tikz.blue}{RGB}{27, 117, 187}
\definecolor{tikz.darkblue}{RGB}{14, 69, 88}
\definecolor{tikz.yellow}{RGB}{251, 175, 63}
\tikzset{
    ->,  
    >=stealth, 
    node distance=1cm, 
    every state/.style={
        inner sep=1pt,
        outer sep=0pt,
        minimum size=0.25cm,
    }, 
    initial text=, 
    font=\scriptsize\ttfamily,
    every initial by arrow/.style={->}
}

\newcommand{\seq}{\mathit{seq}}
\newcommand{\Act}{\mathit{Act}}
\newcommand{\Seq}{\mathit{Seq}}

\title{\LARGE \bf Conversion of LSAT behavioral specifications to automata}
\author{Sander Thuijsman and Michel Reniers
\thanks{*Research leading to these results has received funding from the EU ECSEL Joint Undertaking under grant agreement n\textsuperscript{o} 826452 (project Arrowhead Tools) and from the partners national programs/funding authorities.}
\thanks{Sander Thuijsman and Michel Reniers are with the Department of Mechanical Engineering, Eindhoven University of Technology, Eindhoven, The Netherlands, \texttt{$\lbrace$m.a.reniers, s.b.thuijsman$\rbrace$@tue.nl}}}

\begin{document}

\maketitle
\thispagestyle{empty}
\pagestyle{empty}

\begin{abstract}
The Logistics Specification and Analysis Tool (LSAT) is a model-based engineering tool used for manufacturing system design and analysis.
Using a domain specific language, a system can be specified in LSAT.
In this paper, a conversion method is presented to obtain the system behavior of an LSAT specification in automata structure. 
\end{abstract}

\section{Introduction}
The Logistics Specification and Analysis Tool (LSAT)\footnote{LSAT is developed by ASML, TNO-ESI and Eindhoven University of Technology. The tool and documentation can be found at \url{https://esi.nl/research/output/tools/lsat}.} 
is a workbench for model-based system design and analysis to enable concept flexible design of manufacturing systems.
LSAT enables rapid design-space exploration of supervisory controllers that orchestrate the behavior of these systems.
The foundations of LSAT have been described in literature, without explicitly mentioning the tool itself.
\cite{Basten2020} provides an overview in which a number of LSAT's functions are used; they consider the xCPS flexible manufacturing system \cite{Adyanthaya2017}.
\cite{Basten2020} shows how high-level activities can be used that consist out of low-level actions in order to describe the system behavior in a compact manner.
\cite{Basten2020,Sanden2015,Sanden2016} show how a Finite State Automaton (FSA) can be constructed on activity-level, which is much smaller than the would-be state space on action-level, to express the system behavior.
Requirements are expressed on activity-level, and supervisor synthesis is applied in order to find the \textit{maximally permissive}, \textit{safe}, and \textit{nonblocking} behavior \cite{Cassandras2010}.
Timing characteristics of an activity can be captured using (max,+) algebra \cite{Baccelli1992}.
Using these characteristics, a (max,+) statespace can be constructed.
A throughput or makespan optimal controller of the system can be found in the (max,+) statespace using optimal cycle ratio algorithms \cite{Dasdan2004,Geilen2010}.
Partial-order reduction is applied to improve the computational performance of the (max,+) analysis in \cite{Sanden2018}.
To improve system performance, a critical path technique \cite{Kelley1959} can be applied to identify the system's bottlenecks.
During the engineering process, LSAT has the ability to supply several graphical visualizations, such as Gantt charts, movement plots, probability distributions, and system architecture diagrams.

Except for supervisor synthesis, all mentioned model-based system engineering procedures can be performed using LSAT.
Synthesis is performed using CIF \cite{Beek2014}\footnote{CIF is developed by Eindhoven University of Technology. The tool and documentation can be found at \url{https://cif.se.wtb.tue.nl}.}, which is added to LSAT through a plugin.
CIF is an automata-based language and tool set used for modeling and synthesizing supervisory controllers.

In previous work supervisor synthesis and timing analysis was performed on activity-level \cite{Sanden2015,Sanden2016}.
The abstraction from action-level to activity-level can be beneficial, as performing these computations on action-level can result in scalability issues \cite{Sanden2018thesis}.
However, some computations still require an explicit behavioral model at action-level.
For example, behavioral requirements may be given on action-level.
It is then impossible to (directly) perform supervisor synthesis or formal verification using these requirements without considering the action-level behavior to at least some extent.
In this work we elaborate on the semantics of the specification of a manufacturing system in LSAT, and transform the action-level behavior of such a specification to automata.
Such a model is used to understand the action-level behavior of an LSAT specification and is a prerequisite for developing automata-based supervisor synthesis or verification techniques for these systems.


The paper is structured as follows.
In Section \ref{sec:logisticssys} we explain specification in LSAT, followed by an elaboration on the interacting system components.
Section \ref{sec:behmodel} starts with preliminaries of automata, and then uses the specification elements of Section \ref{sec:logisticssys} to construct sets of automata that describe the behavior of an LSAT specification.
Conclusions are given in Section \ref{sec:conclusions}.


\section{LSAT specification}
\label{sec:logisticssys}
In this section, we first introduce the specification elements that comprise an LSAT specification.
Then, an interpretation of the control system architecture is provided, that uses the conceptual elements of the specification.

\subsection{Specification elements}
\label{sec:elements}
An LSAT specification contains the following elements,
these definitions are strongly based on \cite{Sanden2016}:

\subsubsection{Peripherals}
A set of peripherals $\mathcal{P}$ is defined. 
The set of peripherals is split into two disjoint sets of either \textit{unmovable} $p_u \in \mathcal{P}_u$, or \textit{movable} $p_m\in \mathcal{P}_m$ peripherals.
A movable peripheral $p_m$ contains a set of positions $L_p$.
For an unmovable peripheral no positions are defined.
\subsubsection{Resources}
Each peripheral $p$ is contained within exactly one resource $r \in \mathcal{R}$. 
We assume a function $R:\mathcal{P}\rightarrow\mathcal{R}$, s.t. $R(p)$ is the resource containing $p$.

\subsubsection{Actions}
An unmovable peripheral contains a set of actions $\mathbb{A}_u \subseteq \mathbb{A}$ it can execute.
For every action of an unmovable peripheral the time it takes to execute the action is specified.
This amount of time may be deterministic or stochastic in the form of a normal, triangular, or PERT distribution \cite{Vose2008}.

For a movable peripheral movement actions $\mathbb{A}_m \subseteq \mathbb{A}$ are defined that are directed movements between its locations. 
A movable peripheral can move between all or a subset of its positions.
For the movement actions we assume functions $L_{s}:\mathbb{A}_m \rightarrow {L_p}$ and $L_{t}:\mathbb{A}_m \rightarrow L_p$ that respectively define the source and target location of the movement of the action.
A movement has a speed profile, which is either a second- or third-order profile, and an optional settling time.
Using the speed profile, the time of the movement can be determined \cite{Lambrechts2003}.

An action can be performed on exactly one peripheral. 
We assume a function $A:\mathcal{P}\rightarrow 2^\mathbb{A}$ indicating the actions that are performed by a peripheral.

\subsubsection{Activities}
An activity $\Act \in \mathcal{A}$ is a DAG $(N,\rightarrow)$, consisting of a set nodes $N$ and a set of dependencies $\rightarrow\: \subseteq N\times N$.
We assume a mapping function $M:N\rightarrow (\mathbb{A} \times \mathcal{P}) \cup (\mathcal{R} \times \lbrace rl,cl \rbrace)$, which associates a node to either a pair $(a, p)$ referring to execution of action $a$ on peripheral $p$; or to a pair $(r, v)$ with $v \in \lbrace rl, cl \rbrace$, referring to a claim $(cl)$ or release $(rl)$ of resource $r$. 
Nodes mapped to a pair $(a, p)$ are called action nodes, and nodes mapped to a claim or release of a resource are called claim and release nodes respectively.
All nodes mapped to the same peripheral are sequentially ordered to avoid self concurrency.
All action nodes are preceded by a claim node and succeeded by a release node of the corresponding resource.
Resources are claimed and released not more than once within an activity.
For the corresponding resource, a claim node is always succeeded by a release node, and a release node is always preceded by a claim node.
We assume a function $R:\mathcal{A}\rightarrow 2^\mathcal{R}$, s.t. $R(\Act)$ is the set of resources that $\Act$ uses.

\subsubsection{Dispatching sequences}
We define a finite activity sequence $\omega$ as an ordered list of activities, $\omega=\Act_1;\Act_2;...;\Act_n$.
Sequences can be concatenated using the `$;$' operator; $\omega=\omega_1;\omega_2$.
$\omega_1$ and $\omega_2$ are then called subsequences.
$\varepsilon$ is the empty sequence, so $\omega;\varepsilon=\omega$.
$\omega(i)$ denotes the \textit{i'th} activity in the sequence (in the case of the given sequence that is $Act_i$).
We use notation $\omega(i\!:\!j)$ to denote the subsequence that goes from the \textit{i'th} to the \textit{j'th} activity in the sequence.
$|\omega|$ denotes the length of a sequence.
The set of prefix sequences is given by prefix$(\omega)=\{\varepsilon\}\cup\{\omega(1\!:\!i) | 1 \leq i \leq |\omega| \}$.
We use notation $\omega^n$ to denote $n$ concatenated subsequences $\omega$: $\omega^{n+1}=\omega^n;\omega$, and $\omega^0=\varepsilon$.

A \textit{dispatching sequence} $\seq$ is given as a finite activity sequence, appended by a finite activity sequence that infinitely repeats; $\seq = \omega_1 ; (\omega_2)^\infty$.
Such a structure with a transient phase followed by a periodic phase is consistent with the operational semantics of synchronous data flow graphs \cite{Ghamarian2006}.
LSAT uses these semantics for the timing optimization synthesis algorithms.
Note that $\omega_1$ and $\omega_2$ may or may not be empty, and $\varepsilon^\infty = \varepsilon$.
The prefix operation for a dispatching sequence follows:
prefix$(\omega_1;(\omega_2)^\infty) = $prefix$(\omega_1) \cup \{ \omega_1;({\omega}_2)^n;\omega_p | n\in \mathbb{N}_{\geq 0}, \omega_p\in$prefix$(\omega_2)\}$

A single dispatching sequence can directly be specified, or a set of possible dispatching sequences can be provided by a (network of) FSA.
These FSA have the activities' names as event labels and their synchronous language defines possible dispatching sequences, this is further discussed in Section \ref{sec:combined}.

\subsection{System components}
\label{sec:architecture}
\begin{figure}[ht]
\centering
\includegraphics[scale=0.75]{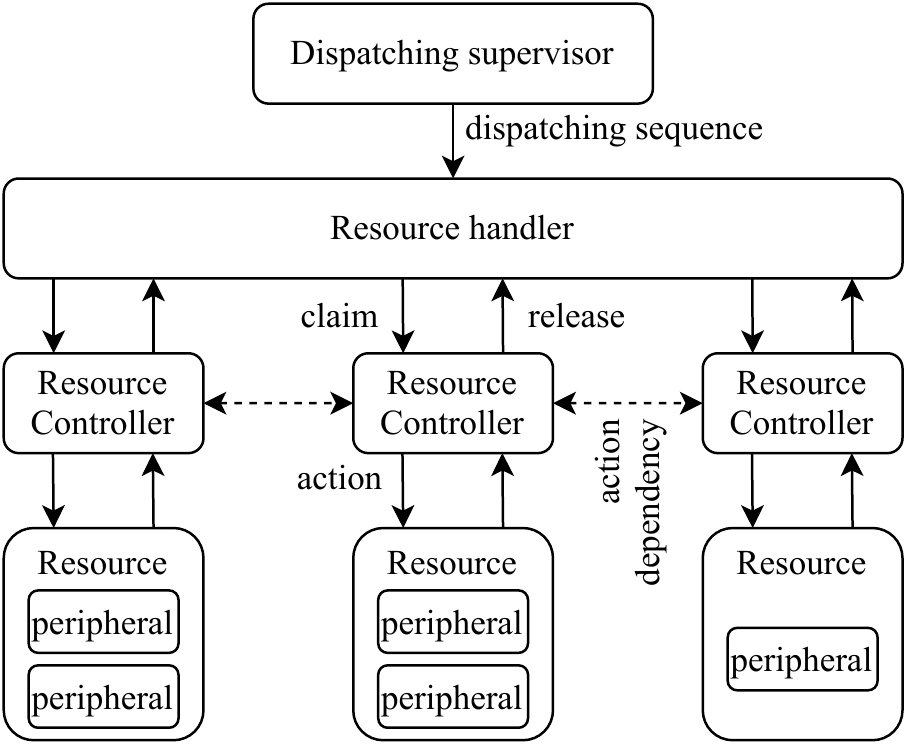}
\caption{Components that model the behavior of an LSAT specification}
\label{fig:architecture}
\end{figure}

The behavior of a manufacturing system specified in LSAT can be modeled by the interacting components that we introduce in this section.
We introduce some vocabulary of these components and their behavior, which we will describe formally in Section \ref{sec:behmodel}. 
Figure \ref{fig:architecture} shows an illustration of the system's components.

The \textit{dispatching supervisor} provides a dispatching sequence to the \textit{resource handler}.
The resource handler acts as a FIFO queue and will claim each resource for the activity that is to use it next in the dispatching sequence, when all incoming dependencies of the particular claim node in the activity are met.
While the resource is claimed, the \textit{resource controller} executes the actions on the peripherals of the resource that are associated with the action nodes in the activity model. 
The resource is released once all incoming dependencies to the release node are met, making the resource available for claiming by a next activity.
There is direct communication between the resource controllers so that dependencies between actions on different resources can (and will) be enforced.

\section{Behavioral model}
\label{sec:behmodel}

In this section we use automata to give a formal behavioral description of the system components introduced in Section \ref{sec:logisticssys}.
In Section \ref{sec:preliminaries} we will provide some preliminaries on automata.
In Section \ref{sec:instantiation} we will discuss instantiation of Activities and the instantiated dispatching sequence.
In the subsections thereafter we will provide definitions for automata describing the different elements of the system.
In Section \ref{sec:combined} the combined behavior of the interacting automata is described.
First, we consider an untimed description of the system behavior; actions occur in an instantaneous and discrete manner in the order as defined by the dependencies between the action nodes in the activities.
An extension of the behavioral description that includes continuous characteristics is discussed in Section \ref{sec:continuous}.

Automata were chosen for this behavioral definition, rather than for example Petri Nets, because there is already existing integrated combined functionality between LSAT and CIF \cite{Beek2014}. 
CIF is an automata-based language and toolset.

\subsection{Preliminaries on Automata}
\label{sec:preliminaries}
The behavioral description might require automata that consist out an infinite amount of states, as shown in Section \ref{sec:instantiation}.
Therefore we use \textit{infinite state automata} in our description, rather than more commonly used finite state automata.
This means that synthesis or verification can not directly be applied to these models.
However, the behavioral semantics are still clear.
The models can act as a baseline to show behavioral equivalence for synthesis or verification techniques that are applied to finite behavioral descriptions.

An infinite state automaton is a \textit{4-tuple} $A=(X,\Sigma,T,X_0)$, where $X$ is the set of states, $\Sigma$ is the set of events, $T\subseteq X\times \Sigma \times X$ is the set of transitions, and $X_0\subseteq X$ is the set of initial states. 
Each of these sets may be finite or infinite \cite{Seatzu2013}.
$\Sigma^*$ is the Kleene-closure of $\Sigma$\cite{Cassandras2010}.
We use notation $A.X$ to refer to states $X$ in automaton $A$.

We use notation $x_{s}\xrightarrow{\sigma}x_{t}$ when there exists a transition $(x_{s},\sigma,x_{t})\in T$.
We extend this notation to strings $w\in \Sigma^*$; $x_{s}\xrightarrow{w\sigma}x_{t}$ if $x_{s}\xrightarrow{w}x_{i}$ and $x_{i}\xrightarrow{\sigma}x_{t}$ for some intermediate state $x_{i}\in X$, and $x \xrightarrow{\varepsilon}x$ for all $x\in X$.
The language of an automaton is given by the set of all possible strings that can be executed from some initial state: $L(A) = \{ w | x_0 \xrightarrow{w} x_{t}, x_0 \in X_0 , x_t\in X\}$.
$\varepsilon$ is the empty string.

The \textit{synchronous composition} \cite{Cassandras2010} of automata $A_1$ and $A_2$ is denoted as $A_1 || A_2$, which creates a new automaton that contains the combined behavior of $A_1$ and $A_2$. 
In this composition, the automata can independently execute their unique events, and shared events are synchronized in lock-step synchronization.
Synchronous composition is associative and commutative, and therefore extends to a set of automata; $\overline{||}(\{A_1, A_2, ..., A_n\})$ denotes $(A_1 || A_2 || ... || A_n)$.

\subsection{Activity instantiation}
\label{sec:instantiation}
\begin{figure}[ht]
\centering
\includegraphics[scale=0.8]{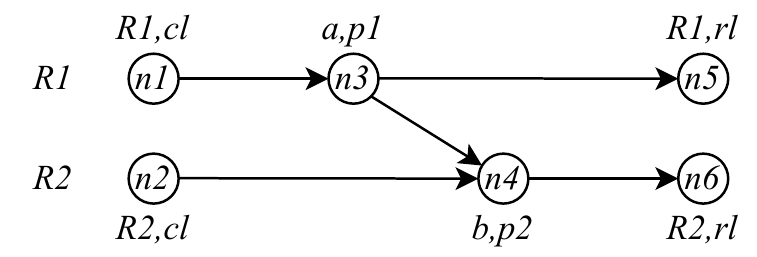}
\caption{Example Activity \textit{Act} DAG}
\label{fig:exampleactivityDAG}
\end{figure}

Two instances of the same activity can be active at the same time.
Let's consider the activity $\Act$ given in Figure \ref{fig:exampleactivityDAG}, and a dispatching sequence $\seq=\Act;(\Act)^\infty$.
Let us refer to the first instance of $\Act$ in this sequence as Activity instance $\Act^1$, and the second instance as $\Act^2$. 
It is possible that $\Act^1$ performs the $a$ action on peripheral $p1$ of $R1$ and releases $R1$, so that $\Act^2$ can claim this resource and perform the $a$ action, before $\Act^1$ has performed the $b$ action.
So, multiple instances of the same activity can be executed concurrently.

The number of concurrent activities of the same instance is unbounded.
Continuing the previous example, it is possible for the activity instance $\Act^n$ to perform the $a$ action before any activity instance $\Act^m$, with $m<n$, has performed action $b$. 
Essentially, the amount of \textit{a}'s that have been performed, has an influence on the amount of \textit{b}'s that still may be performed.
$n$ may be of an arbitrarily large size.
This means that (for some combinations of dispatching sequences and activities) it is required to keep track of an arbitrarily large (unbounded) amount of activity instances.
For each activity instance an automaton is used to track its progress, so there might be infinitely many of such automata.
Also in dispatching these (each time new) activity instances will occur, leading to automata with an infinite amount of transitions and states.
In conclusion, we allow the behavioral description to use an infinite amount of states and/or an infinite amount of automata.

\subsection{Used components}
To construct the system behavior, we will make models of the above descriptions that are relevant to a given dispatching sequence.
E.g., peripherals that are not used, or activity instances that do not appear in the instantiated dispatching sequence are also not included in the model.
For $\seq=\omega_1;(\omega_2)^\infty$, we define the set of used activities $\mathcal{A}_\in = \{\Act\in \mathcal{A}|\exists i : \omega_1(i)=\Act \vee \exists j : \omega_2(j)=\Act\}$ used resources $\mathcal{R}_\in = \{r\in \mathcal{R}| r\in R(\Act), \Act\in \mathcal{A}_\in\}$, and used peripherals $\mathcal{P}_\in = \{p\in \mathcal{P} | M(n) = (a,p), n\in N, (N,\rightarrow)\in \mathcal{A}_\in\}$.

For any activity that appears in the $\omega_2$ part of the dispatching sequence we model an infinte amount of activity instances.
For any activity that appears in the $\omega_1$, but not the $\omega_2$, part of the dispatching sequence we model the amount of activity instances as the number of times it appears in the dispatching sequence.
The set of used activity instances is given as $\tilde{\mathcal{A}}_\in = \{ \Act^j | \Act\in \mathcal{A}, j\in \mathbb{N}_{>0}, 1\leq j\leq |\{i\in \mathbb{N}_{>0}| \omega_1(i)=Act \}| \} \cup \lbrace \Act^j | \Act\in \mathcal{A}, \exists(i\in \mathbb{N}_{>0}) : \omega_2(i)=Act, j\in \mathbb{N}_{>0} \rbrace$.

To denote an instance of $\Act$, without the explicit instance number, we write $\tilde{\Act}$.

\subsection{Resource handler}
\label{sec:resourcehandler}
The resource handler allows a resource to be claimed by one activity instance at a time, and claims the resources for the activity instances as the respective activities appear in the dispatching sequence.
We capture these two functionalities in two seperate sets of automata: \textit{availability automata} and \textit{claiming automata}.

\subsubsection{Availability automata}
Each resource can only be claimed by one activity instance at a time; claim and release actions for this resource must occur in alternating manner.
To model this behavior, we construct a \textit{resource availability automaton} $A_r$ for each resource $r\in \mathcal{R}_\in$ as follows: $A_r=(X,\Sigma,T,X_0)$, where:
\begin{itemize}
\item $X=\lbrace \mathit{released}, \mathit{claimed} \rbrace,$
\item $\Sigma = \lbrace (\tilde{\Act},cl,r), (\tilde{\Act},rl,r) | \tilde{\Act} \in \tilde{\mathcal{A}}_\in, R(\Act)\ni r\}$
\item $T=\lbrace (\mathit{released},(\tilde{\Act},cl,r),\mathit{claimed}) | \tilde{\Act} \in \tilde{\mathcal{A}}_\in, r\in R(\Act) \rbrace \cup \lbrace (\mathit{claimed},(\tilde{\Act},rl,r),\mathit{released}) |$ $\tilde{\Act} \in \tilde{\mathcal{A}}_\in, r\in R(\Act) \rbrace,$
\item $X_0=\lbrace \mathit{released} \rbrace.$
\end{itemize}
Initially each resource is released. 
It can be claimed by any activity instance, by event $(\tilde{\Act},cl,r)$.
After which, it can only be claimed again after it has been released; $(\tilde{\Act},rl,r)$.
The releasing will occur as defined by the Activity automata, discussed later in Section \ref{sec:resourcecontroller}.

\subsubsection*{Example}
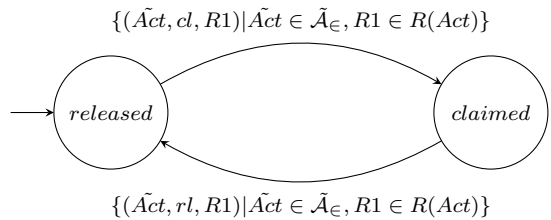
\begin{figure}[ht]
\centering
\begin{tikzpicture}[auto,->,node distance=5cm,align=center,font=\scriptsize\it, uncontrollable/.style={densely dashed}]
\node[state,initial left,initial text ={}, minimum size=1.5cm] (released) {$released$};
\node[state,right of=released, minimum size=1.5cm] (claimed) {$claimed$};

\path[->] (released) edge[bend left] node {$\lbrace (\tilde{Act},cl,R1) | \tilde{Act}\in \mathcal{\tilde{A}}_\in, R1 \in R(Act) \rbrace$} (claimed);
\path[->] (claimed) edge[bend left] node {$\lbrace (\tilde{Act},rl,R1) |\tilde{Act}\in \mathcal{\tilde{A}}_\in, R1 \in R(Act) \rbrace$} (released);
\end{tikzpicture}
\caption{Availability automaton $A_{R1}$ for resource $R1$}
\label{fig:availabilityautomaton}
\end{figure}

An availability automaton for resource $R1$ is given in Figure \ref{fig:availabilityautomaton}.

\subsubsection{Claiming automata}
\label{sec:claimingautomata}
For each resource \textit{r} we can define a \textit{resource claiming automaton} $C_r$, that claims the particular resource in the order that it is used by the activity instances in the dispatching sequence.
We introduce a function $\rho_r$ that reduces a sequence so that it only contains the activity instances that use the particular resource $r$, inductively defined as follows:
\begin{align*}
\rho_r(\varepsilon)     & = \varepsilon\\
\rho_r(\Act;\omega) & = \begin{cases}
\Act;\rho_r(\omega) &\text{if } r\in R(\Act)\\
\rho_r(\omega) &\text{if } r\not\in R(\Act)
\end{cases}
\end{align*}
Given a dispatching sequence $\seq=\omega_1;(\omega_2)^\infty$, we define the resource claiming automaton for resource $r$ as follows: $C_r=(X,\Sigma,T,X_0)$, where:
\begin{itemize}
\item $X=$prefix$(seq_r)$, with $seq_r=\rho_r(\omega_1);(\rho_r({\omega}_2))^\infty$
\item $\Sigma=\lbrace (\tilde{\Act},cl,r) | \tilde{\Act} \in \mathcal{\tilde{A}}_\in, R(\Act)\ni r \rbrace, $
\item $T=\lbrace (\omega_x, (Act^j,cl,r), \omega_x;\Act) | \omega_x \in X, \omega_x;\Act \in X, j=1+|\{i\in \mathbb{N}_{>0} | \omega_x(i)=Act\}| \rbrace $
\item $X_0=\{\varepsilon\}. $
\end{itemize}
The states $X$ are defined by the sequence of activity instances that have been performed on resource $r$.
The set of events $\Sigma$, are the claim actions of all activity instances that use resource $r$ that appear in this sequence.
Each time, the claim action of the first activity in this sequence transitions to the next state.
Initially, all claim actions still need to be performed, so the empty sequence of activities that use resource $r$ is the initial state.
The instance number of the activity instance $(j)$ is incremented by one each time the respective activity appears.
For this, the amount of times the activity appears in the sequence of activities that have been performed thus far in the particular state is used.

\subsubsection*{Example} 
\begin{figure}[ht]
\centering
\subfigure[$C_{R1}$]{
\begin{tikzpicture}[auto,->,node distance=3.5cm,align=center,font=\scriptsize\it, uncontrollable/.style={densely dashed}]
\node[state,initial left, minimum size=1.7cm] (x0) {$\varepsilon$};
\node[state,right of=x0, minimum size=1.7cm] (x1) {$\Act_A^1$};
\node[state,right of=x1, minimum size=1.7cm] (x2) {$\Act_A^1;\Act_B^1$};
\node[state,right of=x2, minimum size=1.7cm] (x3) {$\Act_A^1;\Act_B^1$\\$\Act_B^2$};

\node[right of=x3, node distance=2.8cm, minimum size=1mm] (dot0) {};
\node[state,right of=dot0, node distance=3mm, minimum size=1mm, fill=black] (dot1) {};
\node[state,right of=dot1, node distance=3mm, minimum size=1mm, fill=black] (dot2) {};
\node[state,right of=dot2, node distance=3mm, minimum size=1mm, fill=black] (dot3) {};

\path[->] (x0) edge[] node {$(\Act_A^1,cl,R1)$} (x1);
\path[->] (x1) edge[] node {$(\Act_B^1,cl,R1)$} (x2);
\path[->] (x2) edge[] node {$(\Act_B^2,cl,R1)$} (x3);
\path[->] (x3) edge[] node {$(\Act_B^3,cl,R1)$} (dot0);
\end{tikzpicture}
}
\\
\subfigure[$C_{R2}$]{
\begin{tikzpicture}[auto,->,node distance=3.5cm,align=center,font=\scriptsize\it, uncontrollable/.style={densely dashed}]
\node[state,initial left, minimum size=1.7cm] (x0) {$\varepsilon$};
\node[state,right of=x0, minimum size=1.7cm] (x1) {$\Act_B^1$};
\node[state,right of=x1, minimum size=1.7cm] (x2) {$\Act_B^1;\Act_C^1$};
\node[state,right of=x2, minimum size=1.7cm] (x3) {$\Act_B^1;\Act_C^1$\\$\Act_B^2$};

\node[right of=x3, node distance=2.8cm, minimum size=1mm] (dot0) {};
\node[state,right of=dot0, node distance=3mm, minimum size=1mm, fill=black] (dot1) {};
\node[state,right of=dot1, node distance=3mm, minimum size=1mm, fill=black] (dot2) {};
\node[state,right of=dot2, node distance=3mm, minimum size=1mm, fill=black] (dot3) {};

\path[->] (x0) edge[] node {$(\Act_B^1,cl,R2)$} (x1);
\path[->] (x1) edge[] node {$(\Act_C^1,cl,R2)$} (x2);
\path[->] (x2) edge[] node {$(\Act_B^2,cl,R2)$} (x3);
\path[->] (x3) edge[] node {$(\Act_C^2,cl,R2)$} (dot0);
\end{tikzpicture}
}
\caption{Claiming automata, example}
\label{fig:claimingautomata}
\end{figure}
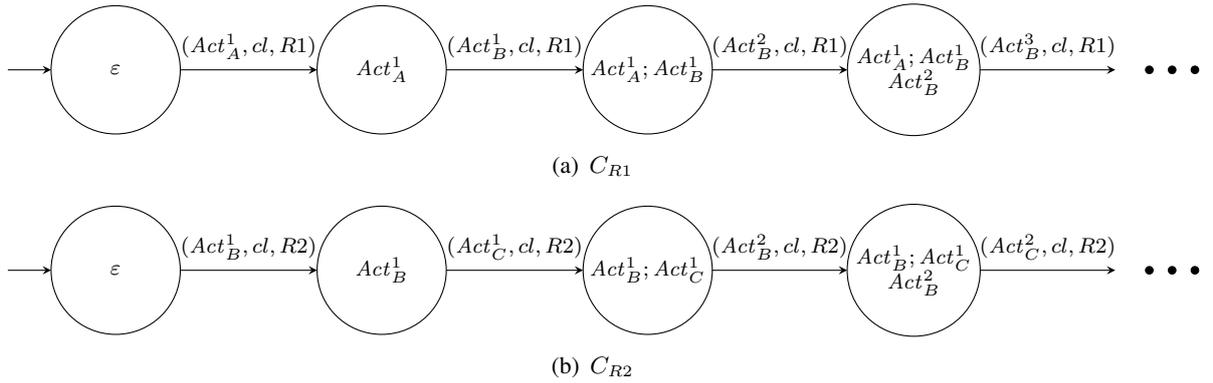

Given dispatching sequence $seq=\Act_A;(\Act_B;\Act_C)^\infty$, where activity $\Act_A$ uses resource $R1$; $Act_B$ uses $R1$ and $R2$; $Act_C$ uses $R2$. 
Applying the definition above provides the two claiming automata shown in Figure \ref{fig:claimingautomata}.

\subsection{Resource controller}
\label{sec:resourcecontroller}
The resource controllers receive claim commands from the resource handler. 
They execute the actions as defined in the respective activity. 
The resource controllers communicate with each other when dependencies between them are defined in the activity definitions.
The behavior of the resource controller can be found by converting the activities (given as DAG) into automata. 

For a given DAG $\Act=(N,\rightarrow)$, we define a \textit{postset} $N'\subseteq N$ as a set of nodes in $N$ that can be reached by iteratively removing nodes and their respecting outgoing dependencies that have no incoming dependencies in $\Act$.
Given $\Act=(N,\rightarrow)$, $N'$ is a \textit{postset} of $N$ if for all $n$ in $N\setminus N'$: whenever $(m,n)\in\rightarrow$, then $m\not\in N'$ (of all nodes that are removed from $N$ to construct $N'$, all their predecessor nodes are removed).
Note that $N$ and $\emptyset$ are postsets of $N$.


The Activity automaton $B_{\tilde{\Act}}$ for activity instance $\tilde{\Act}$ of activity $\Act=(N,\rightarrow)$ can be given as follows:
$B_{\tilde{\Act}}=(X,\Sigma,T,X_0)$, where:
\begin{itemize}
\item $X=\{ N'| N' \text{ is a postset of } N\rbrace,$
\item $\Sigma = \lbrace (\tilde{\Act},a,p) |(a,p)\in N \rbrace,$
\item $T=\{ (N', (\tilde{\Act},a,p),N'\setminus \{n\}) | M(n)=(a,p), N'\in X, n\in N' \},$ 
\item $X_0=\{ N\}.$
\end{itemize}

\subsubsection*{Example} 
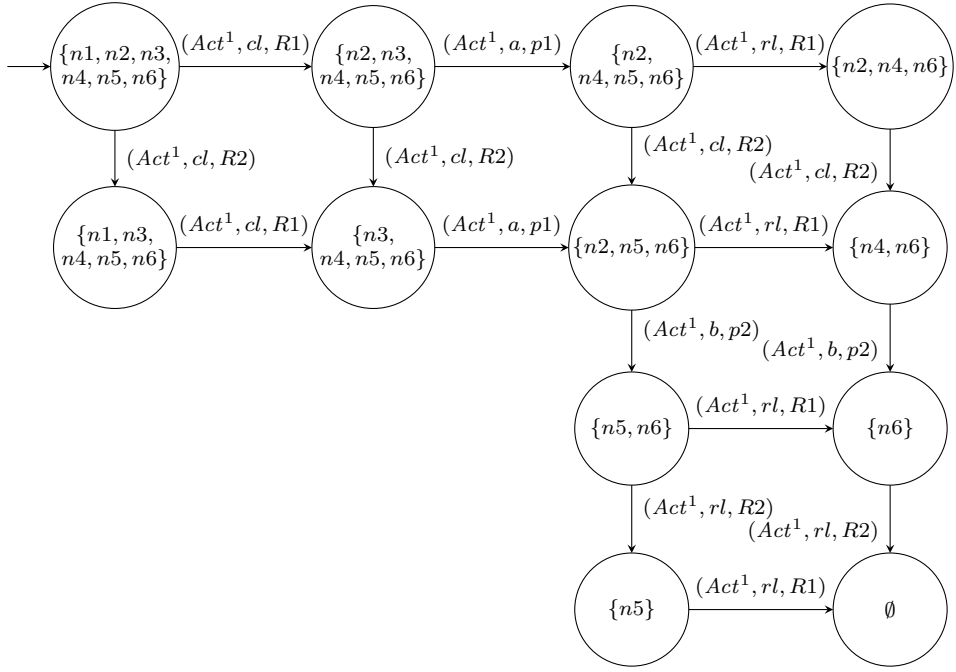
\begin{figure*}[ht]
\begin{center}
\begin{tikzpicture}[auto,->,node distance=3.4cm,align=center,font=\scriptsize\it, uncontrollable/.style={densely dashed}]
\node[state,initial left,initial text ={}, minimum size=1.5cm] (L0L0) {$\{n1,n2,n3,$\\$n4,n5,n6\}$};
\node[state,below of=L0L0, minimum size=1.5cm,node distance=2.4cm] (L0L1) {$\{n1,n3,$\\$n4,n5,n6\}$};

\node[state,right of=L0L0, minimum size=1.5cm,node distance=3.4cm] (L1L0) {$\{n2,n3,$\\$n4,n5,n6\}$};
\node[state,right of=L0L1, minimum size=1.5cm,node distance=3.4cm] (L1L1) {$\{n3,$\\$n4,n5,n6\}$};

\node[state,right of=L1L0, minimum size=1.5cm,node distance=3.4cm] (L2L0) {$\{n2,$\\$n4,n5,n6\}$};
\node[state,right of=L1L1, minimum size=1.5cm,node distance=3.4cm] (L2L1) {$\{n2,n5,n6\}$};
\node[state,below of=L2L1, minimum size=1.5cm,node distance=2.4cm] (L2L2) {$\{n5,n6\}$};
\node[state,below of=L2L2, minimum size=1.5cm,node distance=2.4cm] (L2L3) {$\{n5\}$};

\node[state,right of=L2L0, minimum size=1.5cm,node distance=3.4cm] (L3L0) {$\{n2,n4,n6\}$};
\node[state,right of=L2L1, minimum size=1.5cm,node distance=3.4cm] (L3L1) {$\{n4,n6\}$};
\node[state,right of=L2L2, minimum size=1.5cm,node distance=3.4cm] (L3L2) {$\{n6\}$};

\node[state,below of=L3L2, minimum size=1.5cm,node distance=2.4cm] (L3L3) {$\emptyset$};

\path[->] (L0L0) edge[] node[] {$(\Act^1,cl,R1)$} (L1L0);
\path[->] (L1L0) edge[] node[] {$(\Act^1,a,p1)$} (L2L0);
\path[->] (L2L0) edge[] node[] {$(\Act^1,rl,R1)$} (L3L0);
\path[->] (L0L1) edge[] node[] {$(\Act^1,cl,R1)$} (L1L1);
\path[->] (L1L1) edge[] node[] {$(\Act^1,a,p1)$} (L2L1);
\path[->] (L2L1) edge[] node[] {$(\Act^1,rl,R1)$} (L3L1);
\path[->] (L2L2) edge[] node[] {$(\Act^1,rl,R1)$} (L3L2);
\path[->] (L2L3) edge[] node[] {$(\Act^1,rl,R1)$} (L3L3);

\path[->] (L0L0) edge[] node[] {$(\Act^1,cl,R2)$} (L0L1);
\path[->] (L1L0) edge[] node[] {$(\Act^1,cl,R2)$} (L1L1);
\path[->] (L2L0) edge[] node[yshift=1.5mm] {$(\Act^1,cl,R2)$} (L2L1);
\path[->] (L2L1) edge[] node[yshift=1.5mm] {$(\Act^1,b,p2)$} (L2L2);
\path[->] (L2L2) edge[] node[yshift=1.5mm] {$(\Act^1,rl,R2)$} (L2L3);
\path[->] (L3L0) edge[] node[swap,yshift=-1.5mm] {$(\Act^1,cl,R2)$} (L3L1);
\path[->] (L3L1) edge[] node[swap,yshift=-1.5mm] {$(\Act^1,b,p2)$} (L3L2);
\path[->] (L3L2) edge[] node[swap,yshift=-1.5mm] {$(\Act^1,rl,R2)$} (L3L3);

\end{tikzpicture}
\end{center}
\vspace*{-5mm}
\caption{Example Activity instance automaton for Activity Act}
\label{fig:exampleactivityFSA}
\end{figure*}

In Figure \ref{fig:exampleactivityFSA} the activity automaton for the DAG of Figure \ref{fig:exampleactivityDAG} is shown.

\subsection{Peripherals}
\label{sec:peripherals}
The system contains a set op peripherals $\mathcal{P}$ that execute actions. 
To each peripheral $p$ a state is associated, which is defined by the last action that is executed on the peripheral. 
E.g., for a lamp peripheral, the state of the lamp is turned on if the last executed action was turning on the lamp.
The peripheral automata only contain actions that are used by one of the activities in the dispatching sequence.

A peripheral automaton for an unmovable peripheral $p$ (with $p\in \mathcal{P}_u$) is given as follows. $Q_{p} = (X,\Sigma,T,X_0)$, where:
\begin{itemize}
\item $X=\{ a | \Act\in \mathcal{A}_\in, Act=(N,\rightarrow), n\in N, M(n)=(a,p) \}$
\item $\Sigma = \lbrace (\tilde{\Act},a,p) | \tilde{\Act}\in \tilde{\mathcal{A}}_\in, \Act=(N,\rightarrow), n\in N, M(n)=(a,p) \rbrace,$
\item $T=\lbrace (x_{s},(\tilde{\Act},a,p),a)  | x_{s}\in X, (\tilde{\Act},a,p) \in \Sigma \rbrace,$
\item $X_0=X.$
\end{itemize}

A peripheral automaton for a movable peripheral $p$ (with $p\in \mathcal{P}_m$), is given as follows. $Q_{p} = (X,\Sigma,T,X_0)$, where:
\begin{itemize}
\item $X= \{ \{L_s(a)\} \cup \{L_t(a)\} | \tilde{\Act}\in \tilde{\mathcal{A}}_\in, \Act=(N,\rightarrow), n\in N, M(n)=(a,p)  \}, $
\item $\Sigma = \lbrace (\tilde{\Act},a,p) |  \tilde{\Act}\in \tilde{\mathcal{A}}_\in, \Act=(N,\rightarrow), n\in N, M(n)=(a,p) \rbrace,$
\item $T=\lbrace (L_s(a),(\tilde{\Act},a,p),L_t(a)) | (\tilde{\Act},a,p) \in \Sigma \rbrace,$
\item $X_0=X.$
\end{itemize}
Note that we use the same action events as for the activity automata, so that the peripheral automata and activity automata synchronize.
This means multiple transitions can be added for the same action, as they appear in different activity automata (e.g., $\sigma_1=(\Act^1_A,a_1,p_1), \sigma_2=(\Act^2_A,a_1,p_1)$).
Note that in the movable peripheral definitions sometimes transitions are disabled, these will also not be present in the FSA.
In an LSAT specification, no initial states or positions are specified for the peripherals.
As such, any state of the peripheral automata may be an initial state.

\subsubsection*{Example} 
\begin{figure}[ht]
\begin{center}
\subfigure[Unmovable peripheral automaton]{
\begin{tikzpicture}[auto,->,node distance=2.5cm,align=center,font=\scriptsize\it, uncontrollable/.style={densely dashed}]
\node[state,initial above,initial text ={}, minimum size=0.5cm] (A) {$a$};
\node[state,initial above,initial text ={}, right of=A, minimum size=0.5cm] (B) {$b$};

\path[->] (A) edge[bend left] node {$(\Act_A^1,b,p_u)$} (B);
\path[->] (B) edge[bend left] node {$(\Act_A^1,a,p_u)$} (A);
\path[->] (A) edge[loop left] node {$(\Act_A^1,a,p_u)$} (A);
\path[->] (B) edge[loop right] node {$(\Act_A^1,b,p_u)$} (B);
\end{tikzpicture}
\label{fig:peripheralunmovable}
}
\subfigure[Movable peripheral automaton]{
\begin{tikzpicture}[auto,->,node distance=3cm,align=center,font=\scriptsize\it, uncontrollable/.style={densely dashed}]
\node[state,initial above,initial text ={}, minimum size=1cm] (A) {$left$};
\node[state,initial above,initial text ={}, right of=A, minimum size=1cm] (B) {$middle$};
\node[state,initial above,initial text ={}, right of=B, minimum size=1cm] (C) {$right$};

\path[->] (A) edge[bend left] node {$(\Act_B^1,l\_to\_m,p_m)$} (B);
\path[->] (B) edge[bend left] node {$(\Act_B^1,m\_to\_l,p_m)$} (A);
\path[->] (B) edge[bend left] node {$(\Act_B^1,m\_to\_r,p_m)$} (C);
\path[->] (C) edge[bend left] node {$(\Act_B^1,r\_to\_m,p_m)$} (B);
\end{tikzpicture}
\label{fig:peripheralmovable}
}
\caption{Peripheral automata}
\label{fig:peripheral}
\end{center}
\end{figure}
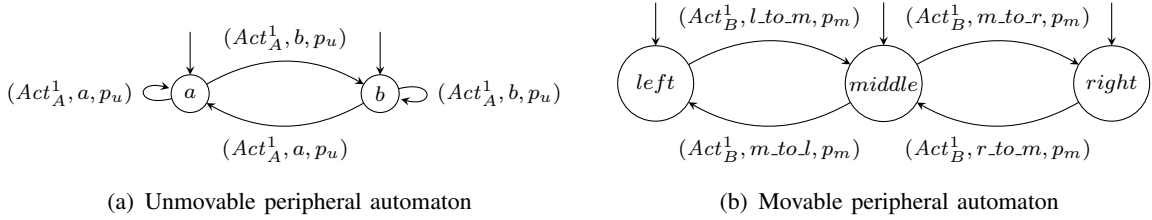

In Figure \ref{fig:peripheralunmovable} an automaton corresponding to an unmovable peripheral $p_u$ on which activity instance $\Act_A^1$ can execute actions $a$ and $b$ is shown.
In Figure \ref{fig:peripheralmovable} an automaton is given that corresponds to a movable peripheral $p_m$ with positions \textit{left, middle,} and \textit{right}, that can move between \textit{left} and \textit{middle}, and between \textit{middle} and \textit{right}.
These actions are used by activity instance $Act_B^1$.

\subsection{Combined system behavior}
\label{sec:combined}
For a given dispatching sequence $\seq$, we can construct the automata described above, their synchronous composition of the relevant components (that appear in $\seq$) describes the system behavior for the sequence in automaton $M^{\seq}$:
\begin{align*}
M^{\seq}=\overline{||}(\{A_r | r\in \mathcal{R_\in} \} \cup \{ C_r | r\in \mathcal{R}_\in \} \cup \{ B_{\tilde{\Act}} | \tilde{\Act} \in \mathcal{\tilde{A}}_\in \} \cup \{ Q_p | p\in \mathcal{P_\in} \})
\end{align*}
As per the definitions above, $M^{seq}$ may be given in a finite or infinite manner, depending whether $seq$ has an infinite part or not.

It is also possible to give a (network of) FSA that describes a set of possible dispatching sequences.
The alphabet of these FSA consists of the activities' names.
We refer to the the synchronous composition of these FSA as FSA $D$.
The language of this automaton is given by a set of words $w\in L(D)$.
Such words can be seen as dispatching sequences, with the activities in the same order in the sequence as the word.
We then write $w \equiv \seq$.
As such, a set of dispatching sequences $\Seq$ is found for the language $L(D)$.
A set of sequences $\Seq$ is considered \textit{complete} for a dispatching FSA $D$ iff: (1) there does not exist a word in the language of $D$ that is not a prefix of any sequence in $\seq$; $\forall w\in L(D) : $ there exists $\seq_p \in$ prefix$(\seq)$ with $\seq\in \Seq$ such that $w\equiv \seq_p$, and (2), there does not exist a prefix of a sequence $\seq$ in $\Seq$ that is not in the language of $D$; $\forall \seq \in \Seq (\forall \seq_p \in$ prefix$(\seq))$: there exists $w\in L(D)$ such that $w\equiv \seq_p$.
For each $\seq\in \Seq$, which may or may not be infinitely many, the automaton $M^{\seq}$ can be constructed as described in the previous sections.
The combined behavior of all possible dispatching sequences is then described in infinite state automaton $M^{\Seq}=(X,\Sigma,T,X_0)$, where:
\begin{itemize}
\item $X = \{ x^{\seq} | x\in M^{\seq}.X, \seq\in \Seq \}$
\item $\Sigma = \{ \sigma | \sigma\in M^{\seq}.\Sigma, \seq\in \Seq \}$
\item $T = \{ (x_s^{\seq},\sigma,x_t^{\seq}) | (x_s,\sigma,x_t)\in M^{\seq}.T, \seq\in \Seq \}$
\item $X_0 = \{ x_0^{\seq} | x_0\in M^{\seq}.X_0, \seq\in \Seq \}$
\end{itemize}
Essentially the automaton $M^{\Seq}$ defines the union of the behavior of all separate automata $M^{\seq}$.
So, $L(M^{\Seq}) = \bigcup_{\seq \in \Seq} L(M^{\seq}) $.

The authors note that complete sets $\Seq$ can be constructed in multiple ways.
For any pair of sets $(\Seq_1,\Seq_2)$, where each set of sequences is complete for automaton $D$, the resulting automaton $M^{\Seq}$ has the same language. Because for any $\seq_1\in \Seq_1$ we know that there exists a $w_1\in L(D)$ such that $\seq_1\equiv w_1$, and there also exists a $\seq_2\in \Seq_2$ such that $\seq_2\equiv w_1$.
Even if $\seq_1$ and $\seq_2$ are different, they represent the same order of activities, and therefore construct the same behavior in $M^{\Seq}$.
An example of different dispatching sequences that construct the same behavior is: $\seq_1=A1;A2;(A1;A2)^\infty$ and $\seq_2=A1;(A2;A1)^\infty$.

\subsection{Continuous behavior}
\label{sec:continuous}
In the above parts automata are constructed that model the behavior of an LSAT specification.
In the conversion from LSAT to the automata, some information was not translated, such as timing and coordinates of the positions of the peripherals.
In this section we shortly discuss how this information can be contained in the automata model.

Hybrid automata\cite{Alur1993} are automata with a set of continuous variables whose values are defined by differential equations.
The differential equation for a continuous variable can be dependent on the state of the automaton, and the continuous variables can be updated (can jump value) when an event occurs.
The peripheral automata described in Section \ref{sec:peripherals} can be modified to hybrid automata with intermediate states that represent an action actively taking place.
A transition leaving these intermediate states can only occur if a certain amount of time has passed, or the target location has been reached.
The time and position are modeled as continuous variables that evolve by a differential equation.
CIF has the functionality to model such hybrid automata.

\section{Conclusions}
\label{sec:conclusions}
LSAT is a model-based engineering tool used for concept flexible manufacturing system design and analysis.
Using a domain specific language, a system can be specified in LSAT.
LSAT has the functionality of analysis, synthesis, visualization and validation that enables to quickly pinpoint improvement areas of the system specification.
An LSAT specification consists out of peripherals, resources, actions, activities and dispatching sequences.
In this paper, the semantics and formalisms of such a specification are elaborated upon.
An action-level behavioral model using automata is given for a system specified in LSAT.
For some LSAT specifications, this behavioral description may require an infinite amount of states and transitions.
This infinite explicit action-level model can act as a baseline in future work, where it can be investigated if verification and synthesis perhaps can be applied to a finite abstraction of this statespace, for action-level behavioral guarantees.

\bibliography{references}

\end{document}